\documentclass[prl,twocolumn,aps,showpacs,nofootinbib,nobibnotes,superscriptaddress,preprintnumbers]{revtex4}
\usepackage{graphicx}
\usepackage{bm}
\usepackage{amsmath}
\usepackage{amssymb}
\usepackage{amsthm}
\usepackage{color}
\usepackage{hyperref}

\newcommand{\diff}{{\rm d}}

\newcommand{\mR}{\mathcal{R}}
\newcommand{\mQ}{\mathcal{Q}}
\newcommand{\mP}{\mathcal{P}}
\newcommand{\divA}{\nabla \cdot A}

\newcommand{\hGamma}{\hat{ \Gamma}}
\newcommand{\hGam}{\hat{ \Gamma}}

\newcommand{\al}{\alpha}
\newcommand{\bt}{\beta}
\newcommand{\ga}{\gamma}

\newcommand{\la}{\lambda}

\newcommand{\be}{\begin{equation}}
\newcommand{\ee}{\end{equation}}

\newcommand{\lp}{\left(}
\newcommand{\rp}{\right)}
\newcommand{\lb}{\left[}
\newcommand{\rb}{\right]}

\newcommand{\mS}{{\mathcal S}}

\newcommand{\bea}{\begin{eqnarray}}
\newcommand{\eea}{\end{eqnarray}}

\newcommand{\lsim}   {\mathrel{\mathop{\kern 0pt \rlap
  {\raise.2ex\hbox{$<$}}}
  \lower.9ex\hbox{\kern-.190em $\sim$}}}
\newcommand{\gsim}   {\mathrel{\mathop{\kern 0pt \rlap
  {\raise.2ex\hbox{$>$}}}
  \lower.9ex\hbox{\kern-.190em $\sim$}}}

\begin{document}
\hspace{5.2in} \mbox{NORDITA-2015-100}\\\vspace{1.53cm} 

\title{Spacetimes with vector distortion: Inflation from generalised Weyl geometry}

\date{\today,~ $ $}

\author{Jose Beltr\'an Jim\'enez}
\email{jose.beltran@cpt.univ-mrs.fr}
\affiliation{CPT, Aix Marseille Universit\'e, UMR 7332, 13288 Marseille,  France.}

\author{Tomi S. Koivisto}
\email{tomi.koivisto@nordita.org}
\affiliation{Nordita, KTH Royal Institute of Technology and Stockholm University, Roslagstullsbacken 23, 10691 Stockholm, Sweden}

\date{\today}

\begin{abstract}

Spacetime with general linear vector distortion is introduced.  
Thus, the torsion and the nonmetricity of the affine connection are assumed to be proportional to a vector field (and not its derivatives).   
The resulting two-parameter family of non-Riemannian geometries generalises the conformal Weyl geometry and some other interesting special cases. 
Taking into account the leading nonlinear correction to the Einstein-Hilbert action results uniquely in the one-parameter extension of the Starobinsky inflation known as the alpha-attractor. The most general quadratic curvature action introduces, in addition to the canonical vector kinetic term, novel ghost-free vector-tensor interactions.  

\end{abstract}
\maketitle

\paragraph{{\bf Spacetime degrees of freedom.}}

In Einstein's General theory of Relativity (GR), gravitation is interpreted as curving of spacetime geometry, and can be described solely in terms of a metric. In addition to a {\it metric} structure however, a manifold representing a physical spacetime must also be endowed with an {\it affine} structure that determines the parallel transport. Though they coincide in GR, a priori these structures are both mathematically and physically independent \cite{Schrodinger:1993:SRZ}. 

Technically this can be formulated simply as the statement that the spacetime connection $\hat{\nabla}$ need not be the Levi-Civita connection $\nabla$ as GR postulates. The $\nabla$ is determined entirely by the metric $g_{\mu\nu}$ as given by the Christoffel symbols,
\be \label{christoffel}
\Gamma^\al_{\bt\ga} = \frac{1}{2}g^{\al\la}\lp g_{\bt\la,\al} + g_{\al\la,\bt} - g_{\al\bt,\ga}\rp\,.
\ee
This is the unique connection that is covariantly conserved, $\nabla_\al g_{\mu\nu}=0$ and symmetric, $\Gamma^\al_{[\bt\ga]}=0$. 
The metric has $D(D+1)/2$ components in a $D$-dimensional spacetime, whereas the connection has $D^3$ components which are, in principle, completely independent degrees of freedom. 
Out of the $D^3$ components, $D^2(D-1)/2$ reside in the antisymmetric part 
\be \label{Gt}
T^\al{}_{\bt\ga} \equiv \hat{\Gamma}^\al_{[\bt\ga]}\,,
\ee
which is called {\it torsion}. The remaining $D^2(D+1)/2$ degrees of freedom are encoded in the {\it non-metricity} tensor 
\be \label{Gnm}
Q_{\al\mu\nu} \equiv \hat{\nabla}_\al g_{\mu\nu}\,. 
\ee
The {\it distortion} $\hat{\Gamma}^\al_{\bt\ga}-\Gamma^\al_{\bt\ga}$ of the affine structure is the combined effect of the torsion and the nonmetricity,
\be
\hat{\Gamma}^\al_{\bt\ga} = \Gamma^\al_{\bt\ga}  + K^\al{}_{\bt\ga} + D^\al{}_{\bt\ga}\,,
\ee 
where the {\it contortion} and the {\it deflection} tensors are defined as
\bea
K^\al{}_{\bt\ga}&=&T^\al{}_{\bt\ga}-T_{\bt\ga}{}^\alpha-T_{\ga\bt}{}^\alpha\, ,\label{contortion} \\
D^\al{}_{\bt\ga}&=&\frac12 g^{\alpha\lambda}\left( Q_{\lambda\beta\gamma}-Q_{\beta\gamma\lambda}-Q_{\gamma\beta\lambda}\right) \label{deflection} \,, 
\eea
respectively \cite{RevModPhys.48.393}.

\paragraph{{\bf Generalising Weyl geometry.}}

The profound idea of gauge symmetry was brought forth within a pioneering non-Riemannian extension of the GR framework due to Hermann Weyl \cite{weyl1921raum}. In Weyl's geometry, the metric compatibility condition is abandoned (while maintaining a symmetric connection) in such a way that the nonmetricity $Q_{\mu\alpha\beta}$ of the connection $\hat{\nabla}$ is determined by a vector $A_\mu$ as follows:
\be \label{nonmetricity}
Q_{\mu\alpha\beta} \equiv \hat{\nabla}_{\mu} g_{\alpha\beta} =-2 A_\mu g_{\alpha\beta}\,.
\ee
Thus, a gauge symmetry arises because this relation is invariant under the  (local) conformal transformation of the metric $g_{\mu\nu}\rightarrow e^{2\Lambda(x)}g_{\mu\nu}$ when simultaneously the vector is transformed as $A_\mu\rightarrow A_\mu-\partial_\mu\Lambda(x)$. The connection coefficients of $\hat{\nabla}$ derived from (\ref{nonmetricity}) are
\be
\hGam^\alpha_{\beta\gamma}=\Gamma^\alpha_{\beta\gamma}-\left(A^\alpha g_{\beta\gamma}-2A_{(\beta}\delta^\alpha_{\gamma)}\right)\,,
\label{Wconnection}
\ee
where the first term represents again the Christoffel symbols (\ref{christoffel}) and the expression inside the brackets is the deflection tensor (\ref{deflection}). The theory obtained by writing the Einstein-Hilbert action in Weyl geometry is a trivial extension as the vector field is non-dynamical, and theories defined by nonlinear functions of the Einstein-Hilbert term turn out to be equivalent to the Palatini-$f(\mR)$ models. More general (Gauss-Bonnet -type) curvature terms however can generate new dynamical, ghost-free vector-tensor theories \cite{Jimenez:2014rna}, see also \cite{Haghani:2014zra}. 

In this letter we propose a {\it linear vector distortion} that generalises the Weyl geometry (\ref{Wconnection}). That is, we consider the most general connection that is determined linearly by a vector field $A_\mu$ without derivatives. The distortion is then given by 3 independent terms\footnote{The axial contortion term $b_4\epsilon^\alpha{}_{\beta\gamma\mu}\tilde{A}^\mu$ is excluded because it would require that the field $\tilde{A}_\mu$ was a pseudovector. Let us mention that adding such a piece would not affect our results as the actions considered in this letter would imply $\tilde{A}_\mu=0$.}:
\be \label{vgamma}
\hat{\Gamma}^\al_{\bt\ga} = \Gamma^\al_{\bt\ga}  
-b_1A^\alpha g_{\beta\gamma}+b_2\delta^\alpha_{(\beta} A_{\gamma)}+b_3\delta^\alpha_{[\beta} A_{\gamma]}\,.
\ee
We see that the original Weyl connection (\ref{Wconnection}) is recovered for $b_2=2b_1=2$ and $b_3=0$. One of the parameters in (\ref{vgamma}) can actually be absorbed into the normalization of the vector field, but we will leave the three of them to track the effects of each term in the following. The torsion (\ref{Gt}) and the non-metricity (\ref{Gnm}) tensors for the vector distortion are, respectively, 
\begin{align}
&Q_{\mu\alpha\beta}  =  (b_3-b_2)A_\mu g_{\alpha\beta} + (2b_1-b_2-b_3) A_{(\alpha}g_{\beta)\mu}\,, \label{Anm} \\
&T^\alpha{}_{\beta\gamma}  =  b_3\delta^\alpha_{[\beta} A_{\gamma]}
\label{At} \,.
\end{align}
Now $b_1$ and $b_2$ contribute only to deflection, while $b_3$ causes also contortion. The torsion-free limit of this geometry, given by $b_3=0$ but general $b_1$ and $b_2$, has been in fact considered earlier in Ref. \cite{0264-9381-8-9-004} (for other investigations into the nonmetric sector, see e.g. \cite{Heinicke:2005bp,Baekler:2006vw}). Some other special cases are listed in the Table \ref{taabeli}.

\begin{table} 
\begin{center}
  \begin{tabular}{| c | c | c | } 
    \hline
  geometry        &    \#   &  constraints \\ \hline \hline
  General           &  $2$                    &  -      \\ \hline
  Riemann           & $0$                     & $b_1=b_2=b_3=0$  \\ \hline
  Dilation (Weyl)                 & $0$                     &  $2b_1-b_2=b_3=0$       \\ \hline
  Generalised Weyl        & $1$                     & $2b_1-b_2=b_3$   \\    \hline
    No dilation                  & $1$                      & $b_2=b_3$    \\ \hline
  Pure deflection            & $0$                      & $b_2=b_3=0$    \\ \hline
  SVN  \cite{0264-9381-8-9-004}           & $1$                    &  $b_3=0$       \\ \hline
  Polar contortion    &  $0$                    &  $b_1=b_2=b_3$       \\ \hline
  (Axial contortion)    &  $0$                    &  $b_1=b_2=b_3=0$, $b_4 \neq 0$)       \\ \hline
  \end{tabular} 
 \caption{Spacetimes with linear vector distortion. The second column indicates the number of free parameters. \label{taabeli}} 
\end{center} 
\end{table}

Amongst them is, as an example, the Weyl-Cartan spacetime that arises from adding torsion to the Weyl connection (\ref{Wconnection}). A remarkable class of geometries is obtained if we set $b_3=2b_1-b_2$ in ($\ref{Anm}$) so that we also recover the Weyl non-metricity relation given in (\ref{nonmetricity}). In detail, given $b_3=2b_1-b_2$, we have $\nabla_{\mu} g_{\alpha\beta}=2(b_1-b_2)A_\mu g_{\alpha\beta}$, 
which is invariant under the Weyl transformation $g_{\mu\nu}\rightarrow e^{2\Lambda(x)}g_{\mu\nu}$ and $A_\mu\rightarrow A_\mu+\partial_\mu\Lambda(x)/(b_1-b_2)$. This presents a whole family of generalized Weyl geometries where the gauge connection of the conformal covariant derivative carries also torsion, as seen from (\ref{At}). 
Thus, we can introduce the covariant derivative $D_\mu g_{\alpha\beta}\equiv\Big[\partial_\mu-2(b_1-b_2)A_\mu\Big]g_{\alpha\beta}$, in terms of which the connection can be expressed as
\be
\hGamma^\mu_{\alpha\beta}=\frac12 g^{\mu\lambda}\Big(D_\alpha g_{\lambda\beta}+D_\beta g_{\alpha\lambda}-D_\lambda g_{\alpha\beta}\Big)+K^\mu{}_{\alpha\beta}\,,
\ee
the first piece respecting the conformal invariance, but the contortion,
\be
K^\mu{}_{\alpha\beta}=\lp b_2-2b_1 \rp \lp A^\mu g_{\alpha\beta}-\delta_\alpha{}^\mu A_{\beta}\rp
\,,
\ee
in general breaking it, unless $2b_1-b_2=0$ and, hence, the torsion vanishes. The Weyl connection (\ref{Wconnection}) is thus the unique conformally invariant connection, but the invariance of the non-metricity relation can be retained in a more general Weyl-Cartan spacetime given a fixed $b_3$.

Let us return to generic spacetimes described by the connection (\ref{vgamma}). The Riemann curvature it generates is given as
\be
\mR_{\mu\nu\rho}{}^\alpha \equiv\partial_\nu\hGamma^\alpha_{\mu\rho}-\partial_\mu\hGamma^\alpha_{\nu\rho}+\hGamma^{\alpha}_{\nu\lambda}\hGamma^{\lambda}_{\mu\rho}-\hGamma^{\alpha}_{\mu\lambda}\hGamma^{\nu}_{\nu\rho}\,,
\label{eq:defRiemann}
\ee
and the corresponding Ricci curvature is just $\mR_{\mu\rho} \equiv \mR_{\mu\alpha\rho}{}^\alpha$. To form the scalar (Ricci) curvature we finally need also the metric, $\mR \equiv g^{\mu\nu}\mR_{\mu\nu}$. 
We find that two extra terms appear due to the nontrivial vector geometry:
\be 
\mR=R-\bt_1 A^2+\beta_2\divA\,,
\label{expRW}
\ee
with (setting $D=4$ from now on)
\bea
\beta_1 & \equiv&-\frac{3}{4}\Big[4b_1^2-8b_1(b_2+b_3)+(b_2+b_3)^2\Big]\,, \label{alpha} \\
\beta_2 & \equiv&-\frac{3}{2}(2b_1+b_2+b_3) \label{beta}\,.
\eea
Because of the projective invariance of the Ricci scalar (or, in general, of the symmetric part of the Ricci tensor), $b_2$ and $b_3$ only enter in the combination $b_2+b_3$. This is so because such a symmetry implies an invariance under the transformation $\hGamma^\alpha_{\mu\nu}\rightarrow \hGamma^\alpha_{\mu\nu} +\delta^\alpha_\mu\xi_\nu$, for an arbitrary vector $\xi_\nu$. This implies that the terms $b_2$ and $b_3$ will give degenerate effects unless the underlying gravitational theory breaks the projective invariance.

\paragraph{{\bf $f(\mR)$ actions.}}

From the result (\ref{expRW}), we see that the pure Einstein-Hilbert action $\mathcal{L} = M_{pl}^2\sqrt{-g}\mR/2$ in a spacetime with the linear vector distortion is equivalent to GR because the last term is a total derivative and the field equations for the vector field\footnote{The mass of the 
vector vanishes if $\bt_1=0$, and becomes tachyonic for all parameter combinations for which $\bt_1<0$. These conditions generalise the result found in the torsion-free case,  \cite{0264-9381-8-9-004}, that in our notation states that the mass is non-tachyonic if $b_1=2(2 - \sqrt{3})b_2 < b_1 < 2(2 + \sqrt{3})b_2$ when $b_3=0$. These conditions can be relevant if one promotes the vector action into the Proca by obtaining the Maxwell term from quadratic curvature invariants as in Ref. \cite{Jimenez:2014rna}.} imply $A_\mu=0$. In order to obtain nontrivial non-Riemannian dynamics, one needs to consider a more general than the pure Einstein-Hilbert form of the action. 

A natural starting point is then to take into account higher order curvature corrections that are expected to become relevant at high energies. 
For this purpose, we will consider prototypical extension of the Einstein-Hilbert action by including an arbitrary dependence upon the Ricci curvature scalar: 
\be
\mS = \frac{M_{pl}^2}{2} \int\diff^4x\sqrt{-g}\mathcal{L}\,, \quad \mathcal{L}=f(\mR)\,. \label{nfr1}
\ee  
It turns out that even in the presence of vector distortion, the actions (\ref{nfr1}) are equivalent to simple scalar-tensor theories. To show this let us first rewrite the lagrangian as
\be \label{nfr2} 
\mathcal{L}= f(\Phi) - \varphi\lp \Phi - R + \bt_1 A^2 - \bt_2 \nabla\cdot A\rp\,.
\ee
It is easy to see that by plugging the constraint $\Phi=\mR$ given by varying with respect to the lagrange multiplier $\varphi$, we recover the original form (\ref{nfr1}). Now let us instead vary with respect to the fields $\Phi$ and $A_\mu$. We obtain, respectively, that
\bea
f'(\Phi)  =  \varphi\,,\label{eomvp} \quad \text{and} \quad
A_\mu  =  -\frac{\beta}{2\alpha\varphi}\partial_\mu\varphi\,.
\eea
Substituting these into the lagrangian (\ref{nfr2}) and dropping the total derivative terms, we obtain the scalar-tensor theory
\be \label{nbd}
\mathcal{L} = \varphi R + \frac{\beta_2^2}{4\beta_1\varphi}\partial_\mu\varphi\partial^\mu\varphi - \lb \varphi\Phi(\varphi)-f\lp\Phi(\varphi)\rp\rb\,.
\ee
The Brans-Dicke coupling parameter is now identified as $\omega_{BD} = -\bt_2^2/4\bt_1$, and in the potential of the scalar the $\Phi(\varphi)$ is solved in terms of the field $\varphi$ from (\ref{eomvp}). 

In the Riemannian limit we have $\omega_{BD} \rightarrow 0$, since (\ref{nfr1}) reduces to the metric $f(R)$ theory. On the other hand, the Palatini-form of $f(\mR)$ gravity corresponds (in $D=4$) to $\omega_{BD}=-3/2$. 
Thus, any theory (\ref{nfr1}) in the vector distorted spacetime is equivalent to metric $f(R)$ theory whenever $\bt_2=0$ and equivalent to Palatini-$f(\mR)$ theory when $\beta_2^2=6\beta_1$.
A subclass of the former are given by our generalized Weyl connection with a conformally invariant metric (in)compatibility condition, i.e., $b_3=2b_1-b_2$.
The non-dilated (or ''dual Weyl'') pure deflector geometry $b_2=b_3=0$ corresponds to $\omega_{BD}=-3/4$. 

We can further transform the theories (\ref{nfr2}) into their Einstein frame $\tilde{g}_{\mu\nu}= \varphi g_{\mu\nu}$. 
In terms of the canonical scalar field 
\be
\tilde{\phi} = \sqrt{\frac{3\al}{2}}M_{pl}\log{\varphi}\,, \quad \text{where} \quad \alpha \equiv  1-\frac{\bt_2^2}{6\bt_1}\,,
\ee
the Einstein frame action then reads
\be \label{efr}
\mS = \int \diff^4x \sqrt{-\tilde{g}}\lb \frac{M_{pl}^2}{2}\tilde{R}-\frac{1}{2}\tilde{g}^{\mu\nu}\tilde{\phi}_{,\mu}\tilde{\phi}_{,\nu} - V(\tilde{\phi})\rb\,,
\ee
where the potential is given by
\be \label{npot}
V(\tilde{\phi}) = \frac{M_{pl}^2}{2\varphi}\lb \Phi(\varphi)-\frac{f\lp \Phi(\varphi)\rp}{\varphi}\rb\,,
\ee
when $\varphi$ is considered as the shorthand for $\varphi \equiv e^{\sqrt{\frac{2}{3\al}}\tilde{\phi}/M_{pl}}$. The action (\ref{efr}) is valid except for Palatini-like theories with $\bt_2=0$, for which the kinetic term of (\ref{efr}) should be erased, the  theory then reducing in vacuum to GR with a cosmological constant.

\paragraph{{\bf Cosmological inflation.}} The new features that arise when the higher curvature terms become dynamically important could have an impact on the phenomenology of inflation in the very early universe. To study this, let us for simplicity study the model defined by taking into account only the leading order quadratic correction to Einstein-Hilbert action. We then need to introduce a mass scale $M$ for the corrections and can write (\ref{nfr1}) as\footnote{The factor of 6 is included so that $M$ is the mass of the scalaron.}
\be
\mS = \frac{M_{pl}^2}{2} \int\diff^4x\sqrt{-g}\mathcal{L}\,, \quad \mathcal{L}= \mR + \frac{\mR^2}{6M^2}\,. \label{staro}
\ee  
In the Riemannian context ($\mR\rightarrow R$), this theory is well known to generate inflation in the early universe, known as the Starobinsky model due to the seminal paper that predicted inflation and nonsingular universes from the leading order curvature corrections to gravity \cite{Starobinsky:1980te}. We now immediately obtain, using (\ref{staro}) in (\ref{nbd}), that in the Jordan frame the theory can be written as
\be
\mathcal{L} = \varphi R + \frac{3}{2}\lp 1-\al\rp\partial_\mu\varphi\partial^\mu\varphi - \frac{3}{2}M^2\lp \varphi - 1\rp^2\,,
\ee
and using (\ref{staro}) in (\ref{efr},\ref{npot}) that the Einstein frame theory is defined by the potential
\be \label{staro_a}
V(\tilde{\phi})=\frac{3}{4}M_{pl}^2 M^2\lp 1 - e^{-\sqrt{\frac{2}{3\al}}\frac{\tilde{\phi}}{M_{pl}}}\rp^2\,.
\ee
This defines a one-parameter generalisation of the original Starobinsky inflationary potential that one recovers in the limit $\al \rightarrow 1$.

The potential (\ref{staro_a}) turns out to represent the so called $\alpha$-attractor parameterisation \cite{Kallosh:2013yoa}. That has been argued to describe generic classes of supergravity-inspired inflationary potentials (wherein the parameter $\al$ is inversely proportional to the curvature of the K\"ahler manifold), and a universal attractor behaviour, reducing effectively to standard chaotic inflation, has been found for various superconformal models in the limit of small $\al$ \cite{Kallosh:2013yoa}. We note also that recently, an ''auxiliary vector field modified gravity'' formulation of the $\al$-attractor model was considered \cite{Ozkan:2015iva}, defined by postulating (\ref{expRW}) with $\bt_1=1$ and $\bt_2 > 0$: our novel geometric framework now provides theoretical underpinnings for such an appearance of a vector field (in a somewhat related approach \cite{Ozkan:2015kma}, the postulate cannot be accommodated). 

The inflationary predictions of the models (\ref{staro_a}) have thus already been computed. One obtains for the scalar spectrum, its tilt and the tensor-to-scalar ratio, respectively,
\bea
\mathcal{P}& = &\frac{2\al}{128\pi^2}\lp\frac{M}{M_{pl}}\rp^2\frac{\lp 1-\varphi\rp^4}{ \varphi^2}\,, \\
n_s & = & 1-\frac{8\varphi^2\lp 1+\varphi\rp}{3\al\lp 1-\varphi^2\rp^2}\,, \\
r &= & \frac{64}{3\al \lp 1-\varphi\rp^2}\,,
\eea
where we recall that $\varphi = e^{\sqrt{2/(3\al)}\tilde{\phi}/M_{pl}}$ in terms of the canonical inflaton $\tilde{\phi}$ in the Einstein frame. In the limit $\al \ll N$, where $N$ is the number of e-folds during inflation, these models share the ''$\alpha$-attractor'' behaviour:
\be
\mathcal{P} \simeq \frac{N^2M^2}{24\pi^2\al M_{pl}^2}\,, \quad
n_s  \simeq  1-\frac{2}{N}\,, \quad
r \simeq \frac{12\al}{N^2}\,,
\ee
generalising the well-known result for Starobinsky inflation with $\al=1$ \cite{1981M,Starobinsky:1983zz}.
In ref. \cite{Ozkan:2015iva}, the equivalent predictions were compared with observations using the latest measurements of temperature and polarisation maps of the cosmic microwave background \cite{Ade:2015lrj}, together with other cosmological data, and the results implied that the present data is unable to distinguish models with $\al \neq 1$. 
The original Starobinsky model fits the data, even with $N=50$ still one sigma, but there the data allows models with more general $\alpha$ too. 
The upper limit we can presently put on $\al$ is of the order $\al \lesssim 100$ if the number of e-folds is considered a free parameter.

\paragraph{{\bf Quadratic actions.}} The Riemann curvature tensor $\mR_{\mu\nu\rho\sigma}$ of the distorted connection $\hat{\Gamma}$, defined in (\ref{eq:defRiemann}), does not have all the symmetries of a purely metric-generated curvature $R_{\mu\nu\rho\sigma}$. This causes that, besides the usual Ricci tensor $\mR_{\mu\nu}=\mR_{\mu\alpha\nu}{}^\alpha$, the co-Ricci tensor 
$\mP_\mu{}^\nu=g^{\alpha\beta}\mR_{\mu\alpha\beta}{}^\nu$ and the homothetic curvature tensor $\mQ_{\mu\nu}=\mR_{\mu\nu\alpha}{}^\alpha$ can be considered independent contractions. The most general quadratic action thus contains several new combinations:
\bea \label{squad}
\mS_{q} & = & \int \diff^D x \sqrt{-g}\Big[\mR^2 + \mR_{\al\bt\gamma\delta}\Big( d_1 \mR^{\al\bt\gamma\delta} + d_2 \mR^{\gamma\delta\al\bt}   \nonumber \\ 
& - & d_3 \mR^{\al\bt\delta\gamma}\Big) -  4\Big( c_1 \mR_{\mu\nu}\mR^{\mu\nu} +  c_2 \mR_{\mu\nu}\mR^{\nu\mu} \nonumber \\
& + & \mP_{\mu\nu}\lp c_3 \mP^{\mu\nu} + c_4 \mP^{\nu\mu} -  c_5 \mR^{\mu\nu} - c_6 \mR^{\nu\mu}\rp  \nonumber \\
& + & c_7 \mQ_{\mu\nu}\mQ^{\mu\nu} + c_8 \mR_{\mu\nu}\mQ^{\mu\nu} + c_9 \mQ_{\mu\nu}\mP^{\mu\nu}\Big)\,   
\Big].
\eea
This action reduces to the (topological) Gauss-Bonnet term in the limit of vanishing distortion, if we set 
\be \label{const}
d_1+d_2+d_3=c_1+c_2+c_3+c_4+c_5+c_6=1\,.
\ee
In general distorted spacetimes, these constraints are not sufficient to guarantee the absence of ghosts instabilities in the theory: rewriting the action in terms of the metric curvature $R_{\al\bt\gamma\delta}$ and the vector field $A_\mu$, we would obtain dangerous coupling terms such as $(\nabla_\mu A^\mu)^2$, $R^{\mu\nu}\nabla_\mu A_\nu$ and $A^2 R$. Generalising the analysis of quadratic theories in Weyl geometry performed in \cite{Jimenez:2014rna} to the extended distorted geometry introduced in this letter, we find that a necessary and sufficient condition for the absence of ghosts in the action (\ref{squad}) with (\ref{const}), is $b_3=2b_1-b_2$. Remarkably, this holds regardless of the coupling parameters $c_i$, $d_i$, as it solely fixes the distortion to correspond precisely to the special one-parameter generalisation of Weyl geometry we have already specified earlier. In this geometry, the quadratic action (\ref{squad}) can be reduced to the interesting vector-tensor theory
\bea \label{squad2}
\mS_{q} & = & \int \diff^D x \sqrt{-g} \Big( R^2 - 4 R_{\mu\nu}R^{\mu\nu}  + R_{\mu\nu\sigma\rho}R^{\mu\nu\sigma\rho}\nonumber \\
&-& \frac{\alpha}{4} F^2 - \bt G^{\mu\nu}A_\mu A_\nu + \xi A^2\nabla\cdot A - \lambda A^4\Big)\, .
\eea
The first line is the Gauss-Bonnet invariant, and in the second line $\alpha$ is some combination of the parameters $c_i$, $d_i$, $b_1$, $b_2$ and the spacetime dimension $D$, while $\beta$, $\xi$ and $\lambda$ only depend on $b_1$ and $D$.
The distorted geometry can thus motivate new viable vector-tensor theories. A detailed study of the physical implications of the theories (\ref{squad2}) will be carried out elsewhere \cite{us}. Let us simply mention that (\ref{squad2}) has isotropic and stable de Sitter fixed points so that $A_\mu$ can be used for dark energy. On the other hand, in \cite{Jimenez:2014rna} it was shown that $A_\mu$ can also be an ideal dark matter candidate. Moreover, the derivative interactions present in (\ref{squad2}) can realize a screening mechanism \`a la Vainshtein for the vector field. We also note that in a recent study \cite{Geng:2015kvs} several interesting possible cosmological applications of Einstein-vector theories belonging to the calls (\ref{squad2}) were suggested (though there without the geometric motivation that was the starting point of this letter). All this proves the rich phenomenology existing for theories formulated within our new geometrical framework.

\paragraph{{\bf Discussion.}}

So far we have discussed exclusively the gravitational sector, but the fundamental issue of matter couplings needs to be addressed in any complete theory of gravitation. In the vector distorted geometry, the Riemannian minimal coupling principle $\partial \rightarrow \nabla$ is naturally generalised to $\partial \rightarrow \hat{\nabla}$. Nevertheless, it turns out that standard matter fields are insensitive to the distortion. This is obvious for standard scalar fields, as they do not couple directly to the connection. Neither does a vector field $V_\mu$, if its  field strength is fundamentally considered as the exterior derivative ${\rm F} ={\rm dV}$. On the other hand, for fermionic fields $\psi$, a connection is required to construct the covariant derivative $D_\mu\psi=(\partial_\mu-\frac14\sigma_{ab}\hGamma^{ab}_\mu)\psi$, where $\hGamma_\mu^{ab}$ is the spin connection and $\sigma_{ab}=[\gamma_a,\gamma_b]/2$ with $\gamma_a$ the gamma matrices. However, an important property of the Dirac lagrangian
\be
\mathcal{L}=\frac{{\rm i}}{2}\left(\bar{\psi}\gamma^a D_a\psi-\gamma^a\bar{D_a\psi}\psi\right)
\ee
is that it generates coupling only to the completely antisymmetric part of the connection. Explicitly, the interaction term ${\rm i}\epsilon_{abcd}\Gamma^{abc}\bar{\psi}\gamma_5\gamma^{d}\psi$ picks up just the axial torsion. Hence, in our set-up, generically matter fields follow the geodesics of the metric Levi-Civita connection and immediate conflicts with the precision tests of equivalence principle are avoided.

In conclusion, the rich geometric structure that emerges by allowing linear vector distortion accommodates consistent and viable theories that can exhibit novel non-Riemannian dynamics if one takes into account higher curvature terms or non-minimal couplings to matter. As the first step we considered curvature-squared actions, obtaining the $\alpha$-attractor generalisation of the Starobinsky inflation and a class of completely new ghost-free vector theories.
These findings encourage further explorations into the physics in spacetimes with vector distortion, to the end of experimentally testing the possible relevance of such geometry in the description of our universe.

\acknowledgments

J.B.J. acknowledges the financial support of A*MIDEX project (n¡
ANR-11-IDEX-0001-02) funded by the ``Investissements d'Avenir" French
Government program, managed by the French National Research Agency
(ANR), MINECO (Spain) projects FIS2011-23000, FIS2014-52837-P and Consolider-Ingenio MULTIDARK
CSD2009-00064. 

\bibliography{refs}

\end{document}